\documentclass[]{aa} 
\usepackage{natbib}
\usepackage{rotating}
\bibpunct{(}{)}{;}{a}{}{,}

\def\cha{{\it Chandra\/  }} 
 
\def\mathnew{\mathsurround=0pt}
\def\simov#1#2{\lower .5pt\vbox{\baselineskip0pt \lineskip-.5pt
\ialign{$\mathnew#1\hfil##\hfil$\crcr#2\crcr\sim\crcr}}}

\def\simless{\mathrel{\mathpalette\simov <}} 

\def\arcsec{\hbox{$^{\prime\prime}$}} 
 
\def\xmm{{\it XMM-Newton\/ }} 
\def\images{{\it images\  }}
\def\MeV{Me\kern-0.11em V} 
\def\keV{ke\kern-0.11em V} 

\usepackage{epsfig, picinpar}
\usepackage{graphics,soul,color}
\usepackage{graphicx, txfonts}
\begin{document}

\title{The Radial Dependence of Temperature and Iron Abundance}
\subtitle{Galaxy Clusters from z=0.14 to z=0.89} 

\author{Steven Ehlert \inst{1,3}
\and M. P.  Ulmer \inst{2,3}}

\institute{Max-Planck-Institut f\"{u}r Kernphysik, PO Box 103980, 69029 
Heidelberg, Germany email: Steven.Ehlert@mpi-hd.mpg.de
\and LAM, P\^ole de l'Etoile Site de Ch\^ateau-Gombert,
38, rue Fr\'ed\'eric Joliot-Curie,
13388 Marseille Cedex 13, France email:m-ulmer2@northwestern.edu
\and Department of Physics and Astronomy, Northwestern University, 2131 Sheridan Road, Evanston, IL 60208-2900, USA }
\authorrunning{Ehlert, S. \& Ulmer, M.P.}

\date{ Received March 18, 2008, Accepted June 6, 2009}

\abstract {The origin and evolution of the intracluster medium
 (ICM) are still not fully understood. A better understanding is not only
interesting in its own right, but it is also important for modeling
hierarchical structure growth and for using cluster
surveys to determine cosmological parameters.}{To determine if there
exists any evidence for evolution in the temperature or iron abundance
gradients between $z \simeq 0.14$ and $z \simeq 0.89$, therefore
elucidating the origin of energy and metal input to the ICM.}{By using
a sample of 35 observations of 31 clusters of galaxies found in the
archival data of \cha and \xmm with redshifts between 0.14 and 0.89,
we derived the temperature and iron abundance radial profiles. To
compare clusters with similar properties, the data were divided into comparable
subsets.}{There is no substantial evidence to suggest that the iron
  abundance radial profiles in galaxy clusters evolve with redshift in any of
the chosen subsets. Temperature radial profiles also do not appear to be
changing with redshift once selection effects are taken into account.}{The lack of evolution
  in the iron profiles is consistent with scenarios where the galaxies in clusters
  are stripped of their gas at higher redshifts. The temperature
  and iron abundance profiles also suggest that the primary source of
  heating in high redshift clusters is the gravitational infall of
  mass. These findings further emphasize the importance of
  modeling the local environment of clusters in cosmological studies
  and have important implications for studies that go to larger redshifts.}

\maketitle

\section{Introduction} Many dynamic processes occur in the high temperature,
iron rich intracluster medium (ICM) of rich clusters of galaxies.
Radiative cooling, supernova driven winds, and ram pressure
stripping are just a few of the processes by which thermal energy and iron may
be injected into the ICM. The theory of
these processes is
discussed in detail elsewhere \citep[e.g.,][]{Sarazin, Rosati,
  Arnaud}. In order to
help resolve which processes most significantly contribute to the
heating and iron enrichment within clusters, the location and evolution with redshift of both iron
abundance and thermal energy can be used to provide significant clues.

Both \cha and \xmm
have the angular resolution to make two-dimensional maps for nearby clusters such
as Coma and Perseus \citep[e.g.,][]{Fabian}, but
the statistical significance of the archival data is not generally sufficient
for these types of measurements in distant clusters. 

Many have already investigated how overall temperatures or iron
abundances might vary with redshift, including \cite{Balestra},
\cite{SZ}, \cite{Tozzi}, and \cite{Sanderson}. So far very little
evidence for the evolution of
the ICM in galaxy clusters up to $ z \simeq 1.0$ has been discovered \citep[]{Matsumoto},
but even subtle hints of changes with redshift could have a profound impact on the
general understanding of the processes involved in the ICM. This current study spatially resolves the
temperature and metal (iron) abundance to higher redshifts than
previous works, as well as
determines whether there exists evidence for changes in the temperature or
iron abundance profile with redshift. 

For higher redshift clusters, studies of the evolution of the radial temperature and
metallicity profiles has been reported out to a redshift $\sim 0.3$ 
\citep{Molendi2008,Leccardi2008}, and here the work
is extended out to $\sim 0.9$. The goal is to search for evolutionary changes in the radial profiles. The
results can be used to determine the relative significance of processes such as ram pressure stripping, Active Galactic Nucleus (AGN) activity, and galactic winds
in clusters of galaxies. The end results can then be compared directly to simulations such as
\cite{Domainko}.  Throughout this paper the concordance model of
cosmology with ${H}_{0}=$71 km s$^{-1}$ Mpc$^{-1}$
$\Lambda_{0}=0.73$ and $\Omega_{\rm{m},0}$=0.27  is assumed.  All
calculations of cosmic distances and times are based on
\cite{Wright2006}.

\section{The Data} 
All data in this study were taken from the \cha and \xmm public
archives, and are listed in Tables
\ref{Table1} \& \ref{Table2}. Most of these observations were
chosen after being included in \cite{SZ}, \cite{Tozzi} or \cite{Molendi2008}. All
clusters listed here that are not found in these studies were found
by searching the \cha \& \xmm archives for galaxy cluster
observations. The resulting sample was chosen to cover a large range of redshift
with reasonable uniformity. This allowed for the sample to be
conveniently divided into subsets which will be compared to one
another, building on previous work at lower redshifts \citep{DeGrandi}.
 The redshifts and coordinates of these
clusters were provided by the NASA/IPAC Extragalactic Database (NED),
and the hydrogen column densities are given by the $n_{H}$ calculator 
provided on the HEASARC website.
Investigations into the element abundances of galaxy clusters suggest that
most of the elemental emission is dominated by iron, and that many other elements have abundance levels that do not change much with redshift \citep{Baumgartner}. This
allowed for the use of iron abundance and metal abundance
interchangeably, and although the model used in this work fit the abundance of all
metals, the results of these fits will hereafter be called the iron
abundance measurements.

\subsection{\cha Data Processing}

All \cha observations used herein were observed with the ACIS camera.  All processing of
\cha data was carried out using CIAO 3.3 and all the contained packages
including Sherpa and ChIPS and followed the general procedure found in
\cite{Tozzi}. For each data set the standard issue EVENTS LEVEL 1 file was
reset to remove all corrections already implemented on the data set. The
process $\it{acis run badpix} \rm$ was run on the reset file to detect all bad
pixels. Next, the standard Level 1 events processing, $\it acis process events
\rm$, was run on the reset file to create a new EVENTS LEVEL 1 file. Detection
of very faint pixels was done whenever the data itself was taken in VFAINT
mode. The new Level 1 file was then filtered to only include the standard event grades 0,2,3,4,
and 6 and also filtered further with the pipeline good time intervals to create a new
EVENTS LEVEL 2 file. The new Level 2 file was processed further using the $\it
destreak \rm$ routine and then using the $\it analyze ltcrv \rm$ routine as part
of the ChIPS package to determine in greater detail the good time intervals.
The bin time used for all data sets was 200 seconds. The de-streaked file was
then filtered again using these good time intervals. As a final step
in processing, the image was then filtered only to include the energy
range of 0.3-10 \keV. Filtering on the energy improves the signal to
noise ratio of the image. The resulting event list was then used in
all subsequent imaging and spectral analysis. There are 25 clusters observed
using 27 \cha observations, with redshifts ranging from 0.142 to 0.890. All of the
cluster data sets, along with their Equatorial coordinates, redshifts,
galactic hydrogen column densities, and net exposure times after all processing are listed in Table
\ref{Table1}. 

\begin{table*}[htbp]
\caption{\label{Table1} Basic information about \cha data sets used
in this study.}
\centering
\begin{tabular}{ c  c  c  c  c  c c }  \multicolumn{7}{c }{
Clusters analyzed from \cha}\\ \hline \multicolumn{1}{c }{Cluster Name}
& \multicolumn{1}{c }{Obs \#} & \multicolumn{1}{c }{RA} &
\multicolumn{1}{c }{DEC} & \multicolumn{1}{c }{\it z } &
\multicolumn{1}{c }{$\rm{n}_{\rm{H}}  (10^{22}\rm{cm}^{-2})$} & \multicolumn{1}{c}{ Exposure Time (s)}\\ \hline\hline Abell 1413 & 1661
& 11 55 18.20 & +23 24 28.80 & 0.142 & 0.0219 & 9749\\ 
Abell 2204 & 499 & 16 32 47.00 & +05 34 33.00 & 0.152 & 0.0567 &
11250\\

Abell 665 & 3586 & 08 30 45.20 & +65 52 55.00 & 0.182 & 0.0431 & 28600\\

RX J0439.0+0520 & 527 & 04 39 02.20 & +05 20 43.00
& 0.208 & 0.1070 & 10830\\ 

Abell 773 & 533 & 09 17 59.40 & +51 42 23.00 & 0.217 & 0.0126 & 10500\\

Abell 697 & 4217 & 08 42 53.30 & +36 20 12.00 & 0.282 & 0.0341 & 18880\\

Abell 611 & 3194 & 08 00 56.90 & +36 03 26.00 & 0.288 &
0.0499 & 35990\\ 

MS 1008.1-1224 & 926 & 10 10 32.33 & +12 39 32.18 & 0.306 & 0.0726 & 42960\\
MS 2137.3-2353 & 928 & 21 40 12.70 & -23 39 27.00 & 0.313 & 0.0355 & 31990\\ 

Abell 1995 & 906 & 14 52 50.40 & +58 02 48.00 & 0.319 & 0.0145 &  43830\\

 ZWCL 1358+6245 &
516 & 13 59 50.60 & +62 31 04.00 & 0.328 & 0.0193 & 45420\\ 

MACS J2228.5+2036 & 3285 & 22 28 34.40 & +20 36 47.00 & 0.412 & 0.0429 & 19700\\
 MACS J2214.9-1359 & 3259 & 22 14 57.30 & -14 00 14.00 & 0.483 &
 0.0328 & 17020\\ 

MACS J1311.0-0310 & 3258 & 13 11 01.70 & -03 10 41.00 & 0.490 & 0.0188 & 15000\\

 MS 0015.9+1609 & 520 & 00 18 33.86 & +16 26 07.75 & 0.541 & 0.0407 & 66940\\ 

MACS 1423.8+2404 & 4195 & 14 23 47.80 & +24 04 41.40 & 0.545 & 0.0238 & 113400\\ 

MS 0451.6-0305 & 529 & 04 54 10.90 & -03 01 07.20 & 0.550 &
0.0500 & 16370 \\

 MS 0451.6-0305 & 902 & 04 54 10.90 & -03 01 07.20 &
0.550 & 0.0500 & 41470\\ 

MACS J2129.4-0741 & 3199 & 21 29 26.20 & -07 41 27.00 & 0.570 & 0.0484 &
17690\\

MS 2053.7-0449 & 551 & 20 56 22.40 & -04 37 43.00 & 0.583 & 0.0462 & 43100\\

MS 2053.7-0449 & 1667 & 20 56 22.40 & -04 37 43.00 & 0.583 & 0.0462 & 43900\\

MACS J0647.7+7015 & 3196 & 06 47 50.20 & +70 14 55.00 & 0.584 & 0.0563
& 18850\\ 

CL 1120+4318 & 5771 & 11 20 07.60 & +43 18 07.00 & 0.600 & 0.0208 & 19740\\
MACS J0744.8+3927 & 3197 & 07 44 53.00 & +39 27 26.00 & 0.686 & 0.0568
& 19690\\ 

MS 1137.5+6625 & 536 & 11 40 23.30 & +66 07 09.00 & 0.782 & 0.0121 &
119500\\

RX J1716.9+6708 & 548 & 17 16 52.30 & +67 08 31.20 & 0.813 & 0.0372 &
50350\\ 

CL J1226.9+3332 & 3180 & 12 26 58.20 & +33 32 48.00 & 0.890 & 0.0137 & 29470\\
\hline
\end{tabular}

\end{table*}

\subsection{\xmm Data Processing} 
All \xmm data analyzed in this project used all three EPIC
  cameras: MOS1, MOS2, and the PN camera. The software used was SAS
  version 7.0, the standard \xmm processing software. All data sets
  were processed using the \images script available on the \xmm
  website. The \images script first runs the fundamental SAS
  algorithms cifbuild, odfingest, epchain, and emchain.  It then
  processes the standard event list further to improve
  signal-to-noise. In order to minimize the significance of flare
  events, the \images script allows for the user to input desired
  Good-Time-Interval (GTI) for each data set. In this case the GTI
  intervals were chosen as time windows where the rate was less than
  35 counts per 100 seconds for the MOS cameras, and 40 counts per 100
  seconds for the PN camera. After GTI filtering the \images script
  then cleans for bad pixels based on a separate input script, with
  default bad pixels provided by the \xmm data center. The standard
  bad pixel table provided by the \images script was used
  here. Although the \images script was originally designed to create
  high quality images from all three cameras, the resulting event
  lists are still useful for spectral studies. Unlike the case of the
  \cha data sets, the event lists used for \xmm analysis have not been
  subject to any energy filtering before determining radial
  profiles. However, all spectra for both instruments are restricted
  to the energy range from 0.5-8.0 \keV. As a final restriction, the
  PATTERNS on the events were limited to single, double, triple,
  and quadruple events for the MOS cameras (PATTERN $\leq$ 12) and limited to
  single and double events for the PN camera (PATTERN $\leq$ 4).   Eight \xmm data sets were
  considered in this study.  They are listed along with
  Equatorial coordinates, redshifts, hydrogen column density, and net exposure times after all processing in
  Table \ref{Table2}. 

\begin{table*}[htbp]
\caption{\label{Table2}Basic information about \xmm data sets used in
analysis.}
\centering
\begin{tabular}{c c c c c c c} \multicolumn{7}{c }{ Clusters
Analyzed from \xmm}\\ \hline \multicolumn{1}{ c  }{ Cluster Name} &
\multicolumn{1}{ c  }{ Obs \#} & \multicolumn{1}{c }{ RA} &
\multicolumn{1}{c }{ DEC} & \multicolumn{1}{c }{ $ \it z$} &
\multicolumn{1}{c }{$ \rm{n}_{\rm{H}}$($10^{22}\rm{cm}^{-2}$)} & \multicolumn{1}{c} {Exposure Time (s)} \\ \hline\hline 
Abell 1763 & 0084230901 & 13 35 17.20 & +40 59 58.00 & 0.223 & 0.0082 & 19500/19500/9197\\
Abell 2390 & 0111270101 & 21 53 34.60  & +17 40 11.00 & 0.228 & 0.0619  & 13900/13900/9445\\
Abell 1835 & 0147330201 & 14 01 02.00  & +02 51 32.00 & 0.253 & 0.0204 & 78120/79450/32060 \\
RX J0256.5+0006 & 0056020301 & 02 56 29.51 & +00 05 28.70 & 0.360 & 0.0650 & 19190/19170/7914\\ 
  RX J0318.2-0301 & 0056022201 & 03 18 28.76 & -03 00 46.70 & 0.370 & 0.0505 & 20270/20340/11570 \\
RX J0426.1+1655 & 0056020401 & 04 26 04.20 & +16 55 48.50 & 0.380 & 0.1910 & 18820/18820/8337\\
 RX J1120.1+4318 & 0107860201 & 11 20 00.91 & +43 18 15.10 & 0.600 & 0.0203 & 22140/22150/16240\\ 
CL J1226.9+3332 & 0200340101 & 12 26 58.00 & +33 32 54.00 & 0.890 & 0.0138 & 77750/77780/60660\\
\hline
\hline
\end{tabular}
\parbox{5.in}{Note: The exposure times for the \xmm data sets are given in the following order: MOS1/MOS2/PN}

\end{table*}

\subsection{Scale Lengths: The Core and Counts Radius}\label{Scales}  
In order to average clusters together in a meaningful way, a scale
length was needed. Two scale lengths were calculated for each 
cluster and used to divide the cluster into four annular regions: the
first scale length was 
calculated by fitting the net intensity to a one-dimensional beta fit
and the second by 
dividing the cluster into four regions with roughly equal numbers of
net counts.The outer radii in these regions are labeled $r_1$, $r_2$, $r_3$, and $r_4$ in
Table \ref{Table3}.  These radial profiles were determined using either the surface brightness
(core) or the net counts (counts), and the background region for these
calculations was always
an annular region at least 5\arcsec\ wide. This annular region was usually sufficiently small to
minimize contamination from line-of-sight point sources while also
containing enough counts ($\simeq $ 100-1000) to be statistically
significant($\sim 3-10 \% 
$). In some cases, the background region
needed to be larger to reach this level of significance, but a region with a
significant number of counts ( $\geq 100$) was always used. The profiles
are not strongly sensitive to the number of counts used in the
background regions, which were chosen specifically to be radially symmetric. For each
cluster, both radii have been calculated using annular regions 2.\arcsec5
wide to determine the intensity profile across the cluster.  Both
scale radii were then used to determine regions for extracting spectra and measuring the
average temperature and iron abundance profiles. The only profiles and measurements
shown in the text will use the counts radius, the reasons for which
will be discussed in \S \ref{Differences}. The measured core radius
and counts regions are found in Table \ref{Table3}.

\subsubsection{The Core Radius}
The basis for using a core radius length scale is that cluster dynamics and the
evolution of the ICM could be inter-related
\citep[e.g.,][]{Rosati}, so a one-dimensional beta fit was performed
on each cluster to determine the core radius scale.
The centers were chosen by eye, and consistent with all
attempts to determine the center of emission analytically. The error
for choosing this center was always a factor of 10 smaller than the
calculated core radius. The radial profile of surface brightness was
calculated using the the CIAO command \it{ dmextract,} \rm
and the net counts annuli described in \S \ref{Scales}, and it was
fit to the Sherpa {\it beta1d} model. The {\it
beta1d} mode fits the data to a function of the form

\begin{equation} I(x)=A \times \left[1+\left(\frac{x-x_0}{{r}_{\rm{c}} }\right)^{2}\right]^{-3\beta+1/2}
\end{equation}

 where r$_{\rm{c}} $ is the core radius. The parameter $\beta$ is a measure of the ratio of kinetic
energy in the galaxies moving in the cluster to the thermal energy of
the cluster \citep[e.g.,][]{Rosati}. The radial fits are described by the two parameters $r_{\rm{c}}$ and $\beta$. The offset
from zero ($x_{0}$) was always fixed at zero, and the proportionality
constant normalizes the fit to the particular data relating it to the central density of the cluster.  Radial
profiles are listed in Table \ref{Table3}, with the first eight
observations being from \xmm. More
elaborate fitting profiles like a double-$\beta$ fit \citep{SZ} could
have been used in lieu of a single-$\beta$ fit, but a single-$\beta$
fit is usually all that statistics require
\citep[e.g.,][]{EttoriTozzi2004}. For clusters with high surface
brightness in their center, however, a more appropriate fit may be the more
elaborate model used
by \cite{Kotov}.

\begin{table*}[htbp]\scriptsize
\caption{Radial profile parameters for all
clusters. The \xmm clusters are listed first. }
\label{Table3}
\centering
\begin{tabular}{ c c c c c c c c c c   } \multicolumn{10}{c }{ Radial Profile
Parameters$^{\ast}$}\\ \hline \multicolumn{1}{ c  }{ Cluster Name} & \multicolumn{1}{
c  }{ RA (Center)} & \multicolumn{1}{ c  }{ DEC (Center)} &
\multicolumn{1}{c }{$ \beta$} & \multicolumn{1}{c} {Core Radius
  } & \multicolumn{1}{c}{$r_{\rm{1}}$} &
\multicolumn{1}{c}{$r_{\rm{2}}$} & \multicolumn{1}{c}{$r_{\rm{3}}$} & \multicolumn{1}{c}{$r_{\rm{4}}$} & \multicolumn{1}{c}{$r_{200}$} \\
\hline\hline 

 Abell 1763 &  13:35:18 &  $+40:59:59$ &
$0.95 \pm 0.03$ & $77.6_{-2.6}^{+2.4}$  & 32.5 & 22.5 & 30 & 37.5 & 717.0\\  

 Abell 2390 &  21:53:37 &  $+17:41:40$ &
$0.572 \pm 0.004$ & $23.4 \pm 0.4$ &  22.5 & 22.5 & 30 & 70 & 729.1 \\

 Abell 1835 &  14:01:02 &  $+02:52:40$ &
$0.635 \pm 0.002$ & $18.88 \pm 0.1$ & 12.5 & 12.5 & 17.5 & 57.5 & 549.8\\ 

 RX J0256.5+0006 &  02:56:34 &  $+00:06:25$ &
$0.74_{-0.01}^{+0.30}$ & $42.7_{-0.7}^{+20.1}$ & 22.5 & 15 & 17.5 & 20 & 249.3 \\ 

 RX J0318.2-0301 & 03:18:34 &
$-03:02:59$ & $0.81_{-0.02}^{+0.30}$ & $42.3_{-1.3}^{+17.5}$ & 20  & 15 & 17.5 & 22.5 & 253.8\\  

 RXJ0426.1+1655 & 04:26:08 &  $+16:55:14$ & $0.61 \pm 0.02 $ & $15.9 \pm
 1.3$  & 12.5 & 7.5 & 10 & 25 & 234.9\\

RXJ1120.1+4318 & 11:20:08 &   $+43:18:05$ &
$0.93 \pm 0.07 $ & $40.2\pm 2.9$ & 17.5 & 10 & 15 & 20 &  149.2\\  

CL J1226.9+3332 & 12:26:58 & $+33:32:45$ & $0.85 \pm 0.03$ &
$23.6 \pm 0.9$ & 12.5 & 7.5 & 10 & 22.5 & 146.8\\ \rm

 Abell 1413 & 11:55:18 & $+23:24:13$&  $0.85 \pm 0.05$ & $45.3\pm 3.3$ & 25& 17.5 &25 & 57.5 & 764.4\\  

Abell 2204$^{\dagger}$ & 16:32:47 & $+05:34:27$  & $0.551 \pm 0.003$ &
$7.5 \pm 0.2$ & 10 & 12.5 & 22.5 & 80 & 558.4\\

Abell 665 & 08:30:59 & $+65:50:12$ & $0.63 \pm 0.02$ &$ 36.6 \pm 2.2$ & 27.5 & 20 & 25 & 50 & 893.3\\

RX J0439.0+0520 & 04:39:02 & $+05:20:46$  & $0.58 \pm 0.02$ &$
10.3 \pm 1.4$ & 10 & 12.5 &20 & 50 & 369.4\\

Abell 773 & 09:17:53 & $+51:43:47$ & $1.51 \pm 0.25 $ & $ 93.2 \pm 11.3
$ & 27.5 & 15 & 20 & 45 & 743.7\\ 

 Abell 611 & 08:00:56 & $+36:03:27$ & $0.85 \pm 0.04$ & $28.8 \pm 1.4$ & 15 &12.5 & 17.5 & 55 & 360.5\\ 
 
 Abell 697 & 08:42:53 & $+36:20:12$ & $0.98 \pm 0.06$ & $67.6 \pm 3.8$ & 30 & 20 & 25 & 67.5 & 448.1\\

 MS 1008.1-1224 & 10:10:33 & $-12:39:59$ & $0.69 \pm 0.04$ & $38.9 \pm
 3.6$ & 25 & 20 & 25 & 55 & 328.2\\ 

 MS 2137.3-2353 & 21:40:16 & $-23:39:43$ & $0.70 \pm 0.01$ &
$9.8 \pm 0.3 $ & 7.5 & 7.5 & 15 & 70 & 270.5\\

Abell 1995 & 14:52:50 & $+58:02:48$ & $1.43 \pm 0.15$ & $69.1 \pm 5.5$  & 25 & 15 & 20 & 50 & 382.0\\

 ZWCL 1358+6245 & 13:59:51 & $+62:31:04$ & $0.57 \pm 0.01$ & $13.4  \pm
 0.8$ & 12.5 & 15 & 20 &42.5 & 289.5\\ 
 
 MACS J2228.5+2036 & 22:28:34 & $+20:37:23$ & $1.47 \pm 0.29$ & $50.5
 \pm 7.1$ & 15 & 10 & 12.5 & 27.5 & 278.5\\

MACS J2214.9-1359 & 22:14:58 & $-14:00:15$ &  $1.00 \pm 0.13$ & $ 36.6
\pm 5.1 $ & 15 & 12.5 & 17.5 & 35 & 244.2 \\ 

 MACS J1311.0-0310 & 13:11:01 & $-03:10:31$& $1.13\pm 0.20$ & $23.7 \pm
 4.4 $ & 10 & 10 & 10 & 37.5 & 201.7\\ 
 
 MS 0015.9+1609 & 00:18:33 & $+16:26:07$ & $1.08 \pm 0.09 $ & $54.2 \pm
 4.0$ & 22.5 & 15 &20 & 57.5 & 221.0\\ 
 
 MACS 1423.8+2404$^{\dagger}$ & 14:23:49 & $+24:04:35$ & $0.62  \pm 0.01$ &
$6.6 \pm 0.3$ & 5 & 7.5 & 17.5 & 47.5 & 162.5\\ 

 MS 0451.6-0305(\#529) &  04:54:12 & $-03:00:58$ & $1.18 \pm 0.19$ &
 $51.7 \pm 7.5$ & 17.5 & 12.5 & 15 & 30 & 208.2\\ 

 MS 0451.6-0305(\#902) &  04:54:11 & $-03:00:48 $& $0.83 \pm  0.03 $ &
 $36.1 \pm 1.5$ & 20 &15 & 20 & 60 & 224.6\\ 

MACS J2129.4-0741 & 21:29:27 & $-07:41:26$  & $0.66 \pm 0.04$ &
$18.6 \pm 2.2$ & 12.5 & 10 & 15 & 37.5 & 185.2\\

MS 2053.7-0449(both) & 20:56:22 & $-04:37:47$ & $0.66 \pm 0.05 $ & $18.2
\pm 2.1$ & 12.5 & 10 & 12.5 & 32.5 & 208.6\\

 MACS J0647.7+7015 & 06:47:50 & $+70:14:55  $& $0.91 \pm 0.08$ &
 $29.8\pm 3.1$ & 15 & 10 & 15 &37.5 & 233.8 \\

 RX J1120.1+4318 & 11:20:06 & $+43:18:06$ & $1.17 \pm 0.31 $ & $42.0 \pm
 9.7$ & 15 & 10 & 12.5 & 22.5 & 141.3 \\

MACS J0744.8+3927$^{\dagger}$ & 07:44:52 & $+39:27:29$ & $0.56 \pm 0.03
$ &$8.0 \pm 1.4$ & 7.5 & 7.5 & 12.5 & 35 & 143.0\\ 

 MS 1137.5+6625 & 11:40:23 & $+66:08:20$ & $0.86 \pm 0.07 $ & $18.8 \pm
 2.0$ & 10 & 7.5 &10 & 27.5 & 133.1\\ 

 RX J1716.9+6708 & 17:16:49 & $+67:08:27$ & $0.89 \pm 0.19$ & $24.4 \pm
 5.8$ & 12.5 & 10 & 12.5 & 37.5 & 124.4\\ 

 CL J1226.9+3332 & 12:26:58 &$ +33:32:48$ & $1.29 \pm 0.30 $ & $27.6
 \pm 5.5$ & 10 & 7.5 & 10 & 27.5 & 141.8\\

\hline
\end{tabular}
\parbox{5.in}{$^{\ast}$All measurements of radial regions are given in
arcseconds.$^{\dagger}$ Denotes the two clusters with extremely
small core radii. See \S \ref{Differences} 
}
\end{table*}

 \subsubsection{The Counts Radius} Upon examination of the one-dimensional
beta fits and their reduced $\chi^{2}$ statistics, it appeared as though
the beta model fits derived from Sherpa were not statistically robust. The
reduced $\chi^{2}$ for these fits was often above 2, and although others have
used similar results \citep{Sakelliou}, it was judged necessary to find a second
method used as to of independently verify the profiles. Therefore, a second set of temperature
and iron profiles were calculated with the same set of data. The second
method used was to divide the cluster into radial regions not by a constant length scale, but instead by giving each region a constant number of counts. The furthest extent of
spectral extraction was also determined by the net counts in a 2.\arcsec5 wide region. 
If the net counts next the 2.\arcsec5 wide region out ever dropped below
20 for \cha or 100 for \xmm data, then this defined the edge of the cluster
due to their low signal-to-noise ratio of the regions beyond this point. The
four calculated counts regions for each cluster are listed in Table
\ref{Table3}.  Dividing the cluster in this manner also allows for 
all of the spectra for one cluster to be comparably
significant instead of having a wide disparity in spectral quality. The spectra were extracted in exactly
the same manner as the core radius spectra, which will be
described in detail in \S \ref{spectrasec}. This process of sub-dividing the cluster will
be called the counts profile or counts radius, hereafter. They will also be listed as $r_{\rm{1}}, r_{\rm{2}}, r_{\rm{3}}$, and $r_{\rm{4}}$  as an abbreviation in the tables, particularly in Table \ref{Table3}

\subsubsection{Differences Between the Two
  Profiles}\label{Differences} 
The radial profiles in each cluster are compared in two ways in Tables
\ref{Table3} \& \ref{Table4}. Table \ref{Table3} shows the extents of
each individual counts region as well as the
calculated core radius and virial radius $r_{200}$ in arcseconds. Table \ref{Table4} shows the
average core and counts regions in terms of kpc. 
It was found that there were insignificant differences 
between the two radii on average, except for
the very outer region. 
The outermost counts regions are usually much wider than
2r$_{\rm c}$ (see the averages given in Table \ref{Table4}). 
Therefore, the average temperature and
iron abundance measurements should be strongly correlated between the core
and counts radius, with differences only being expected in the
outermost regions. Also included in Table \ref{Table3} is the
calculated virial radius in arc seconds. The virial radius is
calculated as found in \cite{Jones}, divided by 1.4 to take into
account the difference between cosmologies (${H}_{0} = 50$ km
s$^{-1}$ Mpc$^{-1}$ in \cite{Jones} while $H_{0} = 71$ km s$^{-1}$
Mpc$^{-1}$ here).  The formula itself is given as

\begin{equation}{ r}_{\rm{200}}=2.779 \times \left( \frac{T} {10\, 
\rm{keV}}\right)
\normalsize ^{0.5}\times
(1+z)^{-1.5} \ {h}_{71}^{-1}\, \rm{Mpc}\end{equation}

and the temperature used was the overall temperature measured by the procedure described 
in $\S$ \ref{spectrasec}. This allows for a
direct comparison with other work that uses the virial
radius. Typically a counts region bin corresponds to about 0.1 
r$_{\rm{virial}}$, thus the profiles discussed here extend out on
average to about 
0.4 r$_{\rm{virial}}$. Although the core and counts radius
were both used as scale radii for temperature and iron abundance
profiles, the measurements for the core radius profile have been
omitted from this text. The core radius results have been omitted
because of their weaker statistics as well as the presence of two
very small core radius measurements, noted in Table
\ref{Table4}. The one-dimensional $\beta$-fit also suffers from bias in
the presence of cool core clusters with high densities in the center.
 The radial profiles produced by the two different methods are very strongly
correlated, and therefore the calculated results were not sensitive to
this choice. Using the counts radius 
allows for more robust statistics and systematics.

\begin{table}[htbp]
\caption{\label{Table4}Average outer extent of the core and counts radius
regions in kpc.}
\centering
\begin{tabular}{ c c c c c }  \multicolumn{5}{c } {Outer extent of
region in kpc}\\ \hline \multicolumn{1}{ c  }{ Radius} &
\multicolumn{1}{c }{Region \#1} & \multicolumn{1}{ c  }{ Region \#2} &
\multicolumn{1}{c }{ Region \#3} & \multicolumn{1}{c }{Region \#4 } \\
[.5mm] \hline\hline

Core & 87 & 173 & 260 & 347 \\ Counts & 84 & 177 & 314 & 607 \\ \hline
 
\end{tabular}

\end{table}

\sethlcolor{yellow}

\subsection{Excluded Data}
Each cluster had to satisfy two general conditions before included
in this study. Each cluster first needed to have a sufficient number of counts to
measure a temperature that was significantly ($ > 2\sigma$) above
zero. All of the clusters in this sample have more than an average of
1,000 counts per detector (3,000 for \xmm observations, 1,000 for \cha
observations) within two core radii, and the minimum luminosity of all
clusters from zero to two core radii is $1 \times 10^{44}$ erg \ $\rm{s^{-1}}$. The
second condition was morphological in nature, as it was necessary to
have a reasonable $\beta$-fit to the data. Any observations that
showed obvious visual signs of recent large merger activity were
immediately excluded. Beyond that, several sets were excluded based on the values derived from the $\beta$-fit. In these cases the best
fit core radius was larger than the furthest extent of the radial
profile. These sets often had $\beta$ values that were unrealistically
large as well, usually well above $\beta =2$. Since the fit was
unreliable in both parameters, these clusters were not considered in
this sample. Table \ref{Table6} lists all the observations investigated, but not
included, in this study and why they were not considered. 

\begin{table*}[htbp]
\caption{\label{Table5}Observations investigated and not included in
this study, along with the reasons they were excluded. }
\centering
\begin{tabular}{ c c c c }  \multicolumn{4}{c } {Clusters Not Analyzed}\\ \hline \multicolumn{1}{ c  }{ Cluster Name} &
\multicolumn{1}{c }{Observatory} & \multicolumn{1}{ c  }{Observation ID \#} &
\multicolumn{1}{c }{ Reason for Exclusion} \\
[.5mm] \hline\hline
Abell 2163 & \cha & 1653 & Bad $\beta$-fit \\
Abell 2218 & \cha & 1454 & Bad $\beta$-fit \\
RX J122+4918  & \cha & 1661 & Bad $\beta$-fit\\
MACS J1149.5+2223 & \cha & 1656 & Bad $\beta$-fit \\
Abell 68 & \cha & 3250 & Bad $\beta$-fit \\
RCS J0439-2904 & \cha & 3577 & Not enough counts \\
Abell 370 & \cha & 515 & Bad $\beta$-fit \\
RX J1347.5-1145 & \cha & 3592 & See Section \ref{RXJ1347} \\
RX J1200.8-0328 & \xmm & 0056020701 & Not enough counts\\
WARP J0152.7-1357 & \xmm & 0109540101 & Merger\\
RX J1334.3+5030 & \xmm & 0111160101 & Merger\\
Sharc-2 & \xmm & 0111160201 & Not enough counts\\
RX J1354.3-0222 & \xmm & 0112250101 & Not enough counts\\
RX J1347.5-1145 & \xmm & 0112960101 & See Section \ref{RXJ1347}\\
MS 1208.71+3928 & \xmm & 0112190201 & Not enough counts\\
 \hline
 
\end{tabular}

\end{table*}

\subsubsection{The Unusual Case of RX J1347.5-1145} \label{RXJ1347}
 The very luminous
cluster RX J1347.5-1145 was also investigated using two observations:
\cha observation \#3592 \citep{Allen} and \xmm observation
\#0112960101 \citep{Gitti}. Although it was sufficiently luminous and symmetric to be included here; the temperatures measured between these
two observations did not agree within $3\sigma$. After this work
began, \cite{Ota1347} found a hot bubble in the southeast region of this
cluster. Thus, it was judged that this cluster is not relaxed 
enough to be considered in
this study and was also excluded.

\subsection{The Spectra} \label{spectrasec}
Spectra for each data set were extracted with
routines in CIAO for \cha or SAS for \xmm. If the data set was from
\cha then the spectra were extracted using the $\it specextract\rm$ 
routine. If the data set was from \xmm, then the OGIP Spectral Products routine
found in the Graphical Interface of SAS after running \it{XMMSelect} \rm 
was used. Spectra were
taken of all four annular regions in both scale lengths. 
When determining the average cluster temperature and iron
abundance a circular
region from $0-2r_{\rm{c}}$ region was used. Background regions were always circles with radii
comparable or larger than two core radii  outside of
the detectable emission and always on the same chip as the cluster
image. The background regions remained consistent within each data
set. These background regions are specifically chosen to contain as
many counts as possible without including sources. Since radial
symmetry is not a major concern with the spectral background, while the
total number of counts is, this
region is not the same background region used to determine the radial
profiles. The software used to do the spectral fitting was XSPEC 12.4.0,
and the MEKAL model in XSPEC was always used along with the TBABS
galactic absorption model. Since the spectra often had low numbers of
counts, the modified Cash statistic was always used to determine the
best-fit temperature and iron abundance simultaneously. The modified
Cash statistic is  ideal for fitting spectra with a low number
of counts in each bin \citep{Cashstat} and also allows for the use of
a local background spectrum instead of fitting the background to a
model. The only 
remaining free parameter in the fit was the normalization. The other 
parameters of the model were fixed: the redshifts were frozen at the
values given by the NASA/IPAC Extragalactic Database (NED) while the
hydrogen column densities were frozen to the values given by the
$n_{H}$ calculator on the HEASARC website. The solar abundance values
used were those from \cite{Anders}. For the \xmm
spectra, all three instruments were fit simultaneously to the same
spectral model to ensure the best use of the available statistics. 
The measured temperature and iron abundance profiles for each cluster are found in Table \ref{Table6}.  
For the four clusters in this study with
multiple observations, the temperature and iron abundance were
determined by fitting all of the observations simultaneously.

\begin{table*}[h]\scriptsize
\caption{\label{Table6}Temperature and iron abundance measurements for all clusters by counts
region.\rm }
\centering
\begin{tabular}{ c c  l   l   l   l  l  l  l l  } \hline \multicolumn{10}{c } {
Temperatures and Iron Abundance Measurements}\\ \hline
  \multicolumn{2}{c}{Cluster Information} &
  \multicolumn{4}{c}{Temperatures in keV} & \multicolumn{4}{c}{Iron
    Abundances in $Z_{\bigodot}$} \\ \hline

 \multicolumn{1}{ c  }{ Cluster Name}  & \multicolumn{1}{ c  }{ $ \it z$} &
\multicolumn{1}{c }{Region \# 1} & \multicolumn{1}{c }{Region \#2} & \multicolumn{1}{c }{ Region \# 3} &
\multicolumn{1}{c }{Region \# 4} & \multicolumn{1}{c}{Region \#1} &
\multicolumn{1}{c}{Region \#2} & \multicolumn{1}{c}{Region \#3} &
\multicolumn{1}{c}{Region \#4}  \\ \hline\hline

Abell 1413 & 0.142 & $6.0 \pm 0.4$ &
$7.8 \pm 0.7$ & $7.4 \pm 0.7$ &
$5.5 \pm 0.4$  & $0.46 \pm 0.14$ &
$0.41 \pm 0.16$ & $0.49 \pm 0.11$ &
$0.73 \pm 0.14$\\ [.5mm]

Abell 2204  & 0.152 & $3.9 \pm 0.1$ &
$6.7 \pm 0.3$ & $10.2 \pm 0.7$ &
$12.1 \pm 0.8$ & $0.98 \pm 0.09$ &
$0.57 \pm 0.09$ & $0.41 \pm 0.11$ &
$0.25 \pm 0.06$\\ [.5mm]

Abell 665 & 0.182 & $7.7 \pm 0.7$ & $7.9 \pm 0.7$ & $8.5 \pm 0.7
$ & $8.1 \pm 0.6$ &  $0.32 \pm 0.12 $ & $0.39 \pm 0.13 $ & $0.47 \pm
0.14$ & $0.31 \pm 0.10$ \\ [.5mm]

RX J0439.0+0520  & 0.208 & $3.1 \pm 0.2 $ &
$4.5 \pm 0.3$ & $4.8 \pm 0.5$ &
$4.2 \pm 0.5$  & $0.50 \pm 0.19$ &
$0.94 \pm 0.28$ & $0.44 \pm 0.13$ &
$0 +0.32$\\ [.5mm]

Abell 773 & 0.217 & $7.3 \pm 1.0 $ & $9.1 ^{+1.9}_{-1.3}$ &
$9.6^{+1.8}_{-1.1}$ & $8.1_{-1.1}^{+1.3}$ & $0.43 \pm 0.22$ & $0.27 \pm 0.23$ & $0.68 \pm
0.28$ & $0.07 _{-0.07}^{+0.25}$\\ [.5mm]

Abell 1763 & 0.223 & $7.5 \pm 0.4$ & $8.3 \pm 0.4$ & $7.0 \pm
0.3 $ & $6.6 \pm 0.3$ & $0.47 \pm 0.08 $ & $0.30 \pm 0.08$ & $0.30 \pm
0.06$ & $0.26 \pm 0.05$\\ [.5mm]

Abell 2390 & 0.228 & $7.3 \pm 0.2$ & $9.4 \pm 0.3$ & $10.0 \pm 0.5 $ &
$9.7 \pm 0.4 $ & $0.43 \pm 0.05$ & $0.43 \pm 0.06$ & $0.29 \pm
0.06$ & $0.17 \pm 0.05$\\

Abell 1835 & 0.253 & $5.0 \pm 0.1$ & $6.8 \pm 0.1$ & $7.2 \pm 0.1$ &
$7.4 \pm 0.1$ & $0.46 \pm 0.02$ & $0.37 \pm 0.02$ & $0.26 \pm
0.02$ & $0.23 \pm 0.02$\\ [.5mm]

Abell 697 & 0.282 & $9.4 \pm 0.9 $ & $9.6 \pm 1.0$ & $ 9.3 \pm
1.0$ & $ 9.1 \pm 1.0$ & $0.50 \pm 0.16$ & $0.37 \pm 0.15$ & $0.54 \pm 0.11$ & $0.31 \pm 0.09$\\[.5mm] 

Abell 611  & 0.288 & $6.7 \pm 0.4$ &
$7.0 \pm 0.5$ & $7.5 \pm 0.6$ &
$6.5 \pm 0.5$ & $0.45 \pm 0.12$ &
$0.24 \pm 0.10$ & $0.21 \pm 0.11$ &
$0.34 \pm 0.07$\\ [.5mm]

MS 1008.1-1224  & 0.306 & $5.9 \pm 0.5$ &
$5.8 \pm 0.6$ & $7.4 \pm 0.8$ & $
6.5 \pm 0.7$ & $0.50 \pm 0.14$ &
$0.35 \pm 0.10$ & $0.37 \pm 0.10$ & $0.04^{+0.10}_{-0.04}$\\ [.5mm]

MS 2137.3-2353  & 0.313 & $4.1 \pm 0.1$ &
$5.1 \pm 0.2$ & $5.9 \pm 0.3$ &
$5.8 \pm 0.5$ & $0.42 \pm 0.06$ &
$0.32 \pm 0.06$ & $0.32 \pm 0.07$ &
$0.35 \pm 0.06$\\ [.5mm]

Abell 1995 & 0.319 & $8.3 \pm 0.7$ & $9.9 \pm 0.8$ & $8.6 \pm
0.8$ & $8.8 \pm 1.0$  & $0.33 \pm 0.12$ & $0.04_{-0.04}^{+0.11}$ & $0.36 \pm 0.10$ & $0.36 \pm 0.11$ \\[.5mm] 

ZWCL 1358+6245  & 0.328 & $4.5 \pm 0.2$ &
$8.9 \pm 0.7$ & $8.5 \pm 1.2$ &
$9.4 \pm 1.8$ & $0.45 \pm 0.10$ &
$0.42 \pm 0.14$ & $0.27 \pm 0.18$ & $0.69_{-0.38}^{+0.10}$\\ [.5mm]

RX J0256.5+0006  & 0.360 & $5.5 \pm 0.4$ &
$5.4 \pm 0.5 $ & $4.4 \pm 0.4$ &
$3.7 \pm 0.2$ & $0.31 \pm 0.13 $& $0.39 \pm 0.14$ & $0.23 \pm 0.10 $& $0.38 \pm 0.14$\\ [.5mm]

RX J0318.2-0301  & 0.370 & $4.0 \pm 0.3$ &
$5.5 \pm 0.5$ & $5.3 \pm 0.6$ &
$6.3 \pm 0.8$ & $0.55 \pm 0.12$ &
$0.27 \pm 0.14$ & $0.21 \pm 0.14$ & $ 0.29 \pm 0.17$\\ [.5mm]

RX J0426.1+1655  & 0.380 & $4.5 \pm 0.4$ &
$6.4 \pm 1.0$ & $4.9 \pm 0.7 $ &
$5.0 \pm 0.7$ & $0.47 \pm 0.15$ &
$0.06_{-0.06}^{+0.18}  $ & $0.07_{-0.07}^{+0.16} $ &
$0.38 \pm 0.22$\\[.5mm]

MACS J2228.5+2036 & 0.412 & $7.6 \pm 1.3 $ & $9.9 \pm 2.2$ &
$7.9 \pm 1.4$ & $ 6.9 \pm 0.8$ & $0.56 \pm 0.21 $ & $ 0.26 \pm 0.19$ &
$0.29 \pm 0.15$ & $0.52 \pm 0.13$ \\[.5mm]  

MACS J2214.9-1359  & 0.483 & $8.0 \pm 1.2 $&
$9.1 \pm 1.7$ & $12.7_{-2.8}^{+5.3}$ &$
8.4 \pm 1.5$ &  $0.27 \pm 0.21 $&
$0.23 \pm 0.22$ & $0.17_{-0.17}^{+0.20}$ & $ 0.13_{-0.13}^{+0.18}$ \\ [.5mm]

MACS J1311.0-0310  & 0.490 & $4.7 \pm 0.5 $&
$6.8 \pm 0.9$ & $7.4 \pm 1.3$ &$
4.8 \pm 0.6$ & $0.19 \pm 0.15 $&
$1.36_{-0.23}^{+0.51} $ & $0.09_{-0.09}^{+0.20}$ & $ 0.22 \pm 0.17$ \\ [.5mm]

MS 0015.9+1609  & 0.541 & $9.3 \pm 0.8 $&
$9.9 \pm 1.0$ & $10.5_{-0.7}^{+1.5}$ &$
10.3 \pm 1.1$  & $0.61 \pm 0.14$&
$0.22 \pm 0.08$ & $0.09 \pm 0.09$ & $0.16 \pm 0.09$ \\ [.5mm]

MACS 1423.8+2404  & 0.545 & $4.1 \pm 0.1$&
$5.8 \pm 0.2$ & $7.1 \pm 0.3$ &$
6.9 \pm 0.5$   & $0.66 \pm 0.08 $&
$0.48 \pm 0.07$ & $0.34 \pm 0.05$ & $0.47 \pm 0.07$\\ [.5mm]

MS 0451.6-0305  & 0.550 & $9.9 \pm 0.8 $&
$8.9 \pm 0.6$ & $9.0 \pm 0.7$ &$
7.3 \pm 0.6$  & $0.64 \pm 0.14 $&
$0.28 \pm 0.10$ & $0.16 \pm 0.10$ & $0.49 \pm 0.13$ \\ [.5mm]

 MACS J2129.4-0741  & 0.570 & $7.0 \pm 1.1 $&
$7.2 \pm 1.2$ & $7.6 \pm 1.2$ &$
5.2 \pm 0.7$ & $0.45 \pm 0.24$&
$0.33 \pm 0.20 $ & $0.41 \pm 0.23$ &$
0.65 \pm 0.17$ \\ [.5mm]

MS 2053.7-0449 & 0.583 & $4.7 \pm 0.7 $ & $4.5 \pm 0.6$ & $5.6
\pm 1.1 $ & $ 3.4 \pm 0.6$ & $0.32 \pm 0.24$ & $0.23 \pm 0.20$ & $0.26 \pm
0.23$ & $0.59 \pm 0.41$\\ [.5mm]

MACS J0647.7+7015  & 0.584 & $14.9^{+4.5}_{-3.0} $&
$11.7_{-2.2}^{+3.6}$ & $10.7_{-2.7}^{+1.7}$ &$
9.5 \pm 1.6$ & $0.43 \pm 0.39 $&
$0.40 \pm 0.25$ & $0.39 \pm 0.22$ & 0+0.18 \\ [.5mm]
 
RX J1120.1+4318 & 0.600 & $5.4 \pm 0.4 $ &
$5.2 \pm 0.4$ & $5.0 \pm 0.4$ &
$4.9 \pm 0.5$ & $0.54 \pm 0.10$ & $0.51 \pm 0.18$ & $0.26
\pm 0.12$ & $0 + 0.16 $ \\ [.5mm]

MACS J0744.8+3927  & 0.686 & $5.3 \pm 0.6$ &
$8.7 \pm 1.7$ & $9.0 \pm 1.6$ &
$5.9 \pm 0.7$ & $0.97_{-0.17}^{+0.40}$ &
$0.18 \pm 0.23$ & $0.20 \pm 0.18$ & $0.24 \pm 0.13$\\ [.5mm]

MS 1137.5+6625  & 0.782 & $7.0_{-0.7}^{+1.0}$ &
$7.5 \pm 1.0$ & $8.5 \pm 1.4$ &
$5.6 \pm 0.7$ & $0.43 \pm 0.18$ & 0 + 0.26 &
$0.38 \pm 0.21$ & $0.24 \pm 0.13 $\\ [.5mm]

 RX J1716.9+6708  & 0.813 & $10.5_{-2.6}^{+3.2}$ &
$4.6_{-0.5}^{+1.1}$ & $4.3 \pm 0.8$ &
N/A  & $0.37 \pm 0.33$ &
$0.23 \pm 0.21$ & $0.88 \pm 0.27$ & N/A \\ [.5mm]

CL J1226.9+3332  & 0.890 & $12.7 \pm 0.8$ &
$11.5 \pm 0.8$ & $10.9 \pm 0.7$ &
$9.3 \pm 0.5$ & $0.34 \pm 0.11 $ &
$0.29 \pm 0.11 $ & $0.08_{-0.08}^{+0.11}$ & $0.06_{-0.06}^{+0.13}  $\\ [.5mm] \hline
\end{tabular}

\end{table*}

\section{Selection Effects}
In any population study, it is important to take into account selection effects within 
the sample itself. Two selection effects are important in this 
study: overall temperature and cooling time. These selection effects combined with the 
separation by redshift lead to eight different subsets of the sample which were then compared 
to one another. In Table \ref{Table7}, 8 subsets defined by their
temperatures and cooling time are listed along with the number of
clusters in each subset. 

\subsection{Temperature and Luminosity}\label{TempSelection} 
The average temperature of this sample of
clusters follows a trend that needs to be taken into
consideration. There is an inherent
expectation that clusters at higher
redshift should exhibit higher temperatures due to the $L_{x}-T$
relationship \cite[e.g.,][and references
therein]{Ota,Pacaud07, EttoriDeGrandiMolendi}. This is because
more distant (higher redshift) clusters need to be more intrinsically
luminous to be detected than low luminosity clusters.  The
X-ray bolometric luminosity is proportional to temperature as
$L_{\rm{x,bol}} \propto T^{2.5}$ \cite[e.g.,][]{Ota}. Thus on average, the observed clusters
at higher redshift should tend to higher average temperatures. It was
necessary to take into account this selection effect when searching for
variations with redshift. Based on the
available sample, the boundary between high and low temperature clusters was set to 6.8 \keV. 
Setting the boundary to this temperature allows for comparably sized
subsets for high and low temperature clusters, even after the two
other selection effects (cooling time and redshift) are taken into
account. The average radial profiles presented below are not strongly dependent
on small changes ($\simeq$ 0.2 keV) to this boundary. 

\subsection{Central Cooling}\label{CoolingSelection} An important mechanism in galaxy cluster
evolution is the process of radiative cooling. The cooling time for a cluster
can be well approximated by \cite{Sarazin} as 

\begin{equation}
 {t}_{\rm{cool}} =
8.5 \times 10^{10} \, \rm{yr} \left (\frac{\it{n}_{ \rm{p} } }  {10^{-3} } \right )^{-1} \left (\frac{\it{T}_{\rm{g}}}{10^{8}
\, \rm{K}}\right )^{1/2}
\end{equation}
 where $n_{p}$ is the particle density in units of $cm^{\rm{-3}}$and $T_{g}$ is the
temperature of the gas in Kelvin. By relating the measured temperature and
luminosity to the density and performing the calculation, the cooling
times for each of these clusters can be calculated. Because high
densities reduce the radiative cooling time, radiative cooling is
potentially significant only in the central regions of a cluster. 
Therefore, all cooling calculations used the temperature and luminosity measured
for the innermost regions. The cooling times were calculated
assuming that the density is constant within the region of zero to
one-half of a core radius and that all of these clusters originally
formed at a redshift of $z=2$.  For the calculation of the density, 
the bolometric luminosity (from $0-100$ keV) and the volume of the $0-0.5 r_{\rm{c}}$ region were used to determine an emissivity for the region, which can then be related to the density by \cite{Sarazin} as

\begin{equation}
{e}^{\rm{ff}} = \left(1.435 \times \rm{10}^{\rm{-27}} \right) \, {T}_{\rm{g}}^{\rm{1/2}} \ {n}_{\rm{p}}^{\rm{2}} \ \rm{erg} \  \rm{cm^{-3}} \ \rm{s^{-1}}
\end{equation}
This assumes that the X-ray emission is all due to thermal
bremsstrahlung and also that the density and temperature of the ICM is constant within $0.5 r_{\rm{c}}$. A single beta fit was used instead of a more
 elaborate fit, and de-projection of
 hydrostatic equilibrium models was not done, as was carried by \cite{SZ}. The cooling
 time only goes as the square root of the temperature, and hence it
 can be seen that any contamination of the temperature determination by the
 outer portions of the cluster (in cooling clusters only) would be a
 weak effect. In comparison with the cooling times calculated in \cite{SZ}, there is only one discrepancy 
 between their
 results and those derived here. The cluster MACS J2129.4-0741 has a cooling time slightly smaller than its age (4.64 Gyr as compared to 4.82 Gyr). Since this difference is
 small compared to the cooling time, and it is not listed as a
 cooling core cluster in \cite{SZ}, it is treated in this study as a
 non-cooling core cluster as well. 
 
The results of these calculations are listed in Table
\ref{Table8}.  The presence of cooling core clusters leads directly to
another important process: central Active Galactic Nucleus (AGN)
activity.  For nearby clusters with cooling cores, it has been shown
that they do not cool as quickly as the theoretical cooling times
suggest \citep{Fabian1994}. 
AGN activity could influence the temperature as far out as $ \simeq 50-100$ kpc, which
is, on average, approximately $0.5r_{\rm c} $ or in the first
counts profile bin. The majority of cool core clusters exhibit the radio emission or bubbles
usually associated with AGN activity \citep{Dunn}.The findings of
\cite{Dunn} also show that only a small fraction of cool core clusters
show no evidence of AGN activity as either bubbles or radio
sources. Therefore, it is expected that central AGN activity is
generally (if not always) present in the cooling core clusters in this
sample as well. For completeness, all clusters were searched for possible counterpart
radio sources within one arc-minute of the cluster position. The
presence of sources was done using the NVSS catalog \citep{Condon},
and the sources were confirmed as being associated with the cluster
using NED. The clusters with likely radio counterparts are denoted
with a dagger in Table \ref{Table8}. The presence of central AGN activity will be assumed in discussing the results for cool core clusters.

\begin{table*}[htbp]
\caption{The eight subsets of the sample used to calculate 
average temperature and iron abundance profiles.  
}
\label{Table7}
\centering
\begin{tabular}{ c c c c c  } \hline \multicolumn{5}{c }
  {Classification of Subsets}\\ \hline \multicolumn{1}{ c  }{ Subset
   \#}  & \multicolumn{1}{c }{Redshift} & \multicolumn{1}{ c  }
{Temperature} & \multicolumn{1}{c }{Cooling}
& \multicolumn{1}{c }{Number} \\ [.5mm] \hline\hline

1 & z$<$0.4 & $<$ 6.8 keV & Yes & 6/6\\
2 & z$<$0.4 & $\geq$ 6.8 keV & Yes & 4/4\\
3 & z$<$0.4 & $<$ 6.8 keV & No & 3/3\\
4 & z$<$0.4 & $\geq$ 6.8 keV & No & 4/4\\

5 & z$>$0.4 & $<$ 6.8 keV & Yes & 2/2\\
6 & z$>$0.4 & $\geq$ 6.8 keV & Yes & 2/2\\
7 & z$>$0.4 & $<$ 6.8 keV & No & 5/3\\
8 & z$>$0.4 & $\geq$ 6.8 keV & No & 9/7\\ 

 \hline
  \end{tabular}
 \parbox{5.in}{The two values in the number column represent the number of observations and the number of unique clusters in that sample, respectively.}

\end{table*}

\begin{table*}[h]\scriptsize
\caption{\label{Table8}  General information for each cluster: the overall (defined here to be from $0-2r_{\rm{c}}$) temperature and iron abundance, the overall luminosity, the total number of counts in the cluster, the calculations relevant to the cooling time, and finally the subset as defined in Table \ref{Table7}.   \rm}
\centering
\begin{tabular}{c c c c c c c c c c c } \hline \multicolumn{6}{c } { Temperatures and
Iron abundances for each full cluster} & \multicolumn{4}{c}{Cooling
    calculations for each central region} & \multicolumn{1}{c}{Analysis Subset}\\ \hline \multicolumn{1}{ c  }{
Cluster Name}  & \multicolumn{1}{ c  }{ $ \it
z$} & \multicolumn{1}{c }{ Temperature} & \multicolumn{1}{c }{Iron
Abundance} & \multicolumn{1}{c}{\# of Counts} &
  \multicolumn{1}{c}{$\mathcal{L}_{0-2}$} &
  \multicolumn{1}{c}{$\mathcal{L}_{0-0.5}$}
  & \multicolumn{1}{c}{ Proton Density } &
  \multicolumn{1}{c}{Cooling Time} & \multicolumn{1}{c}{Age} & \multicolumn{1}{c}{Subset \#} \\ \hline\hline

Abell 1413$^{\dagger}$  & 0.142 & $7.0 \pm 0.3$ &
$0.46 \pm 0.08$ & 11164 & 5.11 & 1.95 & 0.019 & 3.79 & 8.55 & 2 \\

Abell 2204$^{\dagger}$  & 0.152 & $4.3 \pm 0.1$ &
$0.86 \pm 0.08$ & 16382 & 5.17 & 1.09 & 0.22 & 0.24 & 8.44  & 1\\ 

Abell 665$^{\dagger}$ & 0.182 & $8.0 \pm 0.4$ & $0.40 \pm
0.08$ & 12526 & 3.89 & 0.87 & 0.012 & 6.91 & 8.11 & 2\\

RX J0439.0+0520$^{\dagger}$  & 0.208 & $3.6 \pm 0.2$ &
$0.76 \pm 0.17$ & 2018 & 1.96 & 0.75 & 0.08 & 0.57 &
 7.83 & 1\\ 

Abell 773 & 0.217 & $8.0 \pm 0.7$ & $0.32 \pm 0.13$ &
8351 & 9.24 & 5.24 & 0.0058 & 14.72 & 7.74 & 4\\

Abell 1763 & 0.223 & $7.9 \pm 0.2$ & $0.34 \pm 0.04$ &
48845 & 8.81  & 3.52 & 0.0062 & 12.93 & 7.68 & 4\\

Abell 2390$^{\dagger}$ & 0.228 & $8.5 \pm 0.2$ & $0.43 \pm
0.04$ & 36705 & 10.70 & 3.41 & 0.038 & 1.94 &
7.63 & 2\\

Abell 1835$^{\dagger}$ & 0.253 & $6.0 \pm 0.1$ & $0.39 \pm 0.01$ &
242214 & 18.20 & 7.20 & 0.088 & 0.71 & 7.38 & 1\\ 

Abell 697 & 0.282 & ${9.7 \pm 0.7}$ & ${0.35 \pm 0.08}$ &
13442 & 16.6 & 7.75 & 0.0082 & 11.15 & 7.10 & 4\\ 

Abell 611$^{\dagger}$  & 0.288 & $7.0 \pm 0.3$ &
$0.31 \pm 0.06$ & 15213 & 6.53 & 2.64 & 0.019 & 3.92 & 7.04 & 2 \\ 

MS 1008.1-1224  & 0.306 & $6.3 \pm 0.4$ &
$0.35 \pm 0.08$ & 7257 & 4.18 & 1.28 & 0.0082 & 8.49 & 6.88 & 3\\ 

MS 2137.3-2353$^{\dagger}$  & 0.313 & $4.6 \pm 0.1$ &
$0.35 \pm 0.04$ & 16958 & 8.29 & 2.87 & 0.10 & 0.60 & 6.81 & 1\\ 

Abell 1995 & 0.319 & ${8.9 \pm 0.5}$ & ${0.28 \pm 0.10}$ &
34062 & 15.20 & 5.73 & 0.0062 & 13.58 & 6.73 & 4\\

ZWCL 1358+6245$^{\dagger}$  & 0.328 & $6.1 \pm 0.3$ &
$0.39 \pm 0.08$  & 8208 & 3.37 & 0.94 & 0.035 & 1.65 & 6.68 & 1\\ 

RX J0256.5+0006  & 0.360 & $ 5.0 \pm 0.2$ &
$0.34 \pm 0.06$ & 8608 & 4.49 & 1.36 & 0.0061 & 11.32 & 6.40 & 3\\ 

RX J0318.2-0301  & 0.370 & $5.7 \pm 0.3$ &
$0.28 \pm 0.08$ & 7879 & 3.76 & 1.31 & 0.0064 & 9.16 & 6.32  & 3\\

RX J0426.1+1655$^{\dagger}$  & 0.380 & $5.4 \pm 0.4$ &
$0.19 \pm 0.07$ & 33174 & 1.55 & 0.77 & 0.021 & 2.83 &
6.23  & 1\\

MACS J2228.5+2036 & 0.412 & $7.1 \pm 0.5$ &
$0.33 \pm 0.10 $ & 6199 & 15.29 & 6.24 & 0.0081 & 10.30 & 5.97 & 8\\

MACS J2214.9-1359  & 0.483 & $8.8 \pm 0.8 $&
$0.29 \pm 0.12$ & 3399 & 14.36  & 5.20 & 0.011 & 7.61 & 5.43 & 8\\

MACS J1311.0-0310${\dagger}$  & 0.490 & $6.5 \pm 0.6$&
$0.45 \pm 0.13$ & 2111 & 10.57 & 4.60 & 0.021 & 3.09 & 5.38 & 5\\ 

MS 0015.9+1609  & 0.541 & $9.8 \pm 0.6 $ &
$0.28 \pm 0.06$ & 17837 & 23.57  & 3.28& 0.0041 & 21.99 &5.02 & 8\\ 

MACS 1423.8+2404$^{\dagger}$  & 0.545 & $5.3 \pm 0.1 $&
$0.54 \pm 0.05$ & 17674  & 10.06  & 3.31& 0.12 & 0.50 & 4.99 & 5
\\ 

MS 0451.6-0305  & 0.550 & $9.4 \pm 0.7 $&
$0.40 \pm 0.11$ & 14303 & 22.11 & 7.34 & 0.011 & 8.32 &4.96 & 8\\ 

MACS J2129.4-0741 & 0.570 & $7.0 \pm 1.1 $&
$0.41 \pm 0.23$  & 1755 & 10.63 & 2.59 & 0.018 &4.64 & 4.82 & 6\\

MS 2053.7-0449 & 0.583 & $4.9 \pm 0.5 $ & $0.28 \pm 0.10$ & 1733 &
1.95 & 0.41 & 0.0083 & 7.75 & 4.74 & 7\\

MACS J0647.7+7015  & 0.584 & $11.5_{-2.0}^{+2.8}  $&
$0.19_{-0.19}^{+0.26} $ &2717 & 16.27 & 7.58 & 0.013 & 8.79 &4.73 & 8 \\ 
 
RX J1120.1+4318 & 0.600 & $5.2 \pm 0.4$ &
$0.37 \pm 0.11$ & 10239 & 7.22 & 3.28 & 0.0063 & 11.21 & 4.63 & 7\\ 

MACS J0744.8+3927${\dagger}$  & 0.686 & $6.8 \pm 0.7$ &
$0.51 \pm 0.18$ & 1153 & 9.36  &2.66 & 0.063 & 1.09& 4.11 & 6\\

MS 1137.5+6625  & 0.782 & $7.2 \pm 0.6$ &
$0.26 \pm 0.13$ & 4073 &6.03  & 2.07 & 0.014 & 5.61 & 3.59 & 8\\ 

 RX J1716.9+6708  & 0.813 & $4.6 \pm 0.5$ &
$0.57 \pm 0.23$ & 1649 & 6.16 & 2.21 & 0.0088 & 9.69 & 3.43 & 7\\

CL J1226.9+3332  & 0.890 & $11.1 \pm 0.3$ &$ 0.20
\pm 0.05$ & 29846 & 20.56 & 10.06 & 0.014 & 7.42 & 3.07 & 8\\
 
\hline
\end{tabular}
\parbox{7.in}{ The column $\mathcal{L}_{0-0.5}$ is the luminosity from $0-0.5 r_{\rm{c}}$, and $\mathcal{L}_{0-2}$ is the luminosity from $0-2r_{\rm{c}}$. The luminosities are in $erg-s^{-1}$, the times are in Gyr, and the densities are in units of $cm^{-3}$.  Galaxy clusters with a dagger($\dagger$) next to them have confirmed radio sources likely associated with the cluster.  The temperature and iron abundance have the same units as in Table \ref{Table6}.}
\end{table*}

\section{Results}
After dividing the sample into eight comparable subsets based on temperature, cooling time, and
redshift, relevant average profiles were calculated and compared. These divisions
had to be made so as to make valid comparisons between profiles derived from low and high redshifts. Since there is currently little evidence for
evolution in galaxy clusters even at high redshifts \citep{Matsumoto, Mushotzky1997}, the null
hypothesis for statistical tests will be that cluster abundance and temperature profiles do not change with
redshift. See also \cite{Leccardi2008,Molendi2008} for $z$ = 0.1-0.3 studies. Only if the statistics
conclusively suggest rejecting this hypothesis will evolution be considered.

\subsection{Zero iron abundances and N/A values}
As seen in Table 6, there are several instances where the best fit iron
abundance was zero. As a compromise
between ignoring these points altogether and producing an
un-weighted average, the derived uncertainties to these points were
used as statistical weights (one over the error squared).  Since these zero
values were most likely due to an insufficient number of detected
photons rather than a true lack of iron in the ICM, this averaging was
judged to be a valid approach.  As a cross check, these points were
excluded from the averaging as well, and the results were all within
$1\sigma$ of each other in both weighted and unweighted averages (see \S \ref{Averages}).  Also, there was  one instance where the
measurement failed to converge for both the temperature and iron
abundance simultaneously, and no measurement errors could be
calculated. This measurement is listed with a ``N/A'' in Table \ref{Table6}.

\subsection{Averaging Procedures}\label{Averages} In order to make the most robust study of the data, two averaging
procedures(weighted and unweighted) were done on each subset listed in the following sections. The
primary averaging process used was a weighted average that took into account the measurement error on
each observation.  For a given sample of $N$ clusters, the weighted temperature/iron abundance average
$\bar{M}$ is calculated using the  measurements $m_{i}$ and the measurement error $\sigma_{i}$ as 

\begin{equation}
\bar{M}=\frac{\sum_{i=1}^{N}
  \frac{m_{i}}{\sigma_{i}^{2}}}{\sum_{i=1}^{N}\frac{1}{\sigma_{i}^{2}}}
\pm \left(\sum_{i=1}^{N}\frac{1}{\sigma_{i}^{2}}\right)^{-1/2}
\end{equation}

As a check to ensure that no significant biases arise due to the weighting procedure, a standard
(unweighted) mean and standard deviation on that mean were also calculated, both of which only depend
on the measurements themselves and not the error on those measurements. The unweighted averages further
demonstrate the robustness of these measurements, as the choice of weighted or unweighted averaging
does not affect the basic conclusions. 

\subsection{Average Profiles}
The average temperature and iron abundance profiles are written out in Tables \ref{Table9} \&
\ref{Table10}. They are organized by subset and radial region, and show the unweighted values in
parentheses next to the weighted averages. Since the weighted mean emphasizes measurements with the
smallest errors and there was substantial variance in the quality of the spectra, the weighted mean was
chosen to be the basis for the subsequent figures and discussion. However, both means exhibited the same general trends and results.

\begin{table*}[h]
\caption{Average temperature as a function of counts
radius, separated by redshift, temperature and cooling times. Unweighted averages are
in parentheses.}
\label{Table9}
\centering
\begin{tabular}{ c c c c c  } \hline \multicolumn{5}{c } {Temperatures in
keV}\\ \hline \multicolumn{1}{ c  }{ Subset \#}  & \multicolumn{1}{c }{Region
\#1} & \multicolumn{1}{ c  }{Region \#2} & \multicolumn{1}{c }{ Region \#3}
& \multicolumn{1}{c }{Region \#4 } \\ [.5mm] \hline\hline

1 & $4.67(4.18) \pm 0.04 (0.27)$ & $6.51(6.38)  \pm 0.07(0.63)$ &
 $7.05(6.92) \pm 0.08(0.87)$ & $7.26(7.32) \pm 0.09(1.23)$ \\[.5mm]

2 & $7.06(6.91) \pm 0.16(0.37)$ & $8.37(8.04) \pm 0.24(0.50)$ &
 $8.68(8.36) \pm 0.29(0.60)$ & $7.51(7.45) \pm 0.24(0.92)$\\[.5mm]

3 & $ 4.81(5.15) \pm 0.23(0.57)$ & $5.54(5.55) \pm 0.31(0.11)$ &
 $5.0(5.69) \pm 0.29(0.89)$ & $4.18(5.52) \pm 0.22(0.91)$ \\[.5mm] 

4 & $7.81(8.13) \pm 0.28(0.49)$ & $8.69(9.12) \pm 0.35(0.28)$ &
 $7.34(8.61) \pm 0.24(0.59)$ & $6.90(8.15) \pm 0.24(0.57)$ \\[.5mm]

5 & $4.11(4.38) \pm 0.11(0.30)$ & $5.81(6.26) \pm 0.18(0.49)$ &
 $7.16(7.29) \pm 0.28(0.15)$ & $6.08(5.89) \pm 0.38(1.06)$ \\[.5mm]

6 & $5.65(6.13) \pm 0.52(0.85)$ & $7.68(7.95) \pm 0.96(0.79)$ &
 $8.12(8.27) \pm 0.97(0.68)$ & $5.56(5.55) \pm 0.51(0.31)$ \\[.5mm]  
 
7 & $5.34(6.86) \pm 0.32(1.81)$ & $4.93(4.78) \pm 0.33(0.23)$ &
$4.94(4.97) \pm 0.36(0.38)$ & $4.21(2.75) \pm 0.37(1.44)$ \\[.5mm]  

8 & $9.47(9.89) \pm 0.37(1.09)$ & $9.48(9.78) \pm 0.37(0.55)$ &
$9.83(10.03) \pm 0.40(0.63)$ & $7.90(8.18) \pm 0.29(0.63)$ \\[.5mm]  

 \hline
 \end{tabular}
\end{table*}

\begin{table*}[h]
\caption{Average iron abundance as a function of counts
radius, separated by redshift, temperature and cooling times. Unweighted averages are
in parentheses.
}
\label{Table10}
\centering
\begin{tabular}{ c c c c c  } \hline \multicolumn{5}{c } { Iron
    Abundance in Solar Units}\\ \hline \multicolumn{1}{ c  }{ Subset
    \#}  & \multicolumn{1}{c }{Region
\#1} & \multicolumn{1}{ c  }{Region \#2} & \multicolumn{1}{c }{ Region \#3}
& \multicolumn{1}{c }{Region \#4 } \\ [.5mm] \hline\hline

1 & $0.48(0.55) \pm 0.02 (0.09)$ & $0.37(0.45)  \pm 0.02(0.12)$ &
$0.27(0.30) \pm 0.02(0.05)$ & $0.25(0.32) \pm 0.02(0.09)$ \\[.5mm]

2 & $0.42(0.42) \pm 0.04(0.03)$ & $0.38(0.37) \pm 0.05(0.04)$ &
$0.33(0.37) 
\pm 0.05(0.07)$ & $0.27(0.39) \pm 0.04(0.12)$\\[.5mm]

3 & $ 0.46(0.45) \pm 0.08(0.07)$ & $0.34(0.34) \pm 0.07(0.04)$ &
$0.28(0.27) 
\pm 0.06(0.05)$ & $0.13(0.24) \pm 0.06(0.10)$ \\[.5mm] 

4 & $0.44(0.43) \pm 0.06(0.04)$ & $0.20(0.25) \pm 0.05(0.07)$ &
$0.37(0.47) 
\pm 0.05(0.09)$ & $0.27(0.25) + 0.04(0.06)$ \\[.5mm]

5 & $0.56(0.43) \pm 0.07(0.24)$ & $0.51(0.92) \pm 0.07(0.44)$ &
$0.32(0.22) 
\pm 0.05(0.13)$ & $0.43(0.35) \pm 0.07(0.13)$ \\[.5mm]

6 & $0.66(0.71) \pm 0.19(0.26)$ & $0.26(0.26) \pm 0.15(0.08)$ &
$0.28(0.31) \pm 0.14(0.11)$ & $0.39(0.45) \pm 0.10(0.21)$ \\[.5mm]

7 & $0.50(0.41) \pm 0.09(0.07)$ & $0.34(0.32) \pm 0.11(0.09)$ &
$0.34(0.47) \pm 0.10(0.21)$ & $0.08(0.20) \pm 0.15(0.20)$ \\[.5mm]

8 & $0.47(0.47) \pm 0.06(0.05)$ & $0.25(0.24) \pm 0.05(0.05)$ &
$0.16(0.22) 
\pm 0.05(0.05)$ & $0.22(0.23) \pm 0.05(0.08)$ \\[.5mm]  
 
 \hline
 \end{tabular}
\end{table*}

\subsection{Temperature Profiles}
The temperature profiles of  Table \ref{Table9} suggest that the
overall temperature and cooling time completely define the temperature
profiles of clusters. The average temperature profiles for cooling
core clusters exhibit a clear increase in temperature with radius, while
clusters without cool cores are either isothermal or have temperatures
that decrease with radius. Similar temperature profiles for
non-cool core clusters has been confirmed by \cite{Sanderson} \&
\cite{Molendi2008}. 

There are also no differences between the temperature profiles of
high and low temperature clusters of comparable cooling times with the
exception of an overall offset due to the higher overall
temperatures. Because the temperature profiles do not seem to deviate
from expectations based on their overall temperature and cooling
times, the discussion in \S \ref{disc} will focus primarily on the measured iron abundance
profiles.

\subsection{Iron Profiles} \label{IronProfs} 
The iron abundance radial profiles are now taken into consideration.

\subsubsection{Low Temperature, Cool Core Clusters}
\begin{figure}
\includegraphics[width=\columnwidth]{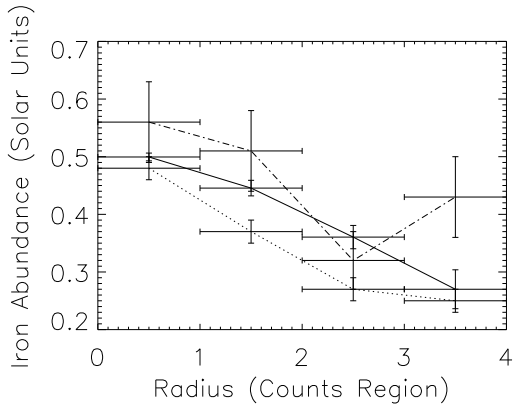}
\caption{\label{Fig1} Iron abundance as a function of radius for
clusters with low temperatures and short cooling times, separated by 
redshift. The dotted line corresponds to subset \# 1 ( $z< 0.4$) while the dashed
line corresponds to subset \# 5 ($z > 0.4$). The solid line is from \cite{DeGrandi}.}

\end{figure}

The first subsets to consider are those of low temperature, cool core
clusters. These clusters have been shown to have strong iron abundance
gradients \citep{DeGrandi} at low redshifts and are often associated
with central AGN activity. Not only do the two subsets considered here
(subsets \# 1 \& 5) show no evidence for evolution with redshift, the
iron abundance profile is almost identical to the profiles of still lower
($z \leq 0.1$) redshift clusters discussed in \cite{DeGrandi}. The results of
\cite{DeGrandi} have been re-projected to the length scale used here and
corrected for the differences in their solar abundance model. All
three profiles are shown in Fig. \ref{Fig1}.  From
Fig. \ref{Fig1} there is no evidence for
evolution in the iron abundance of cool core clusters
from $<z> \simeq 0.7$ to $<z> \simeq 0$. 
It should also be noted in Table \ref{Table10} that low
temperature cool core clusters (subsets 1 \& 5 ) do not have 
statistically significant
differences in overall iron abundance or in average iron abundance
profiles versus redshift. The same lack of difference is 
also seen when comparing high temperature
cool core clusters (subsets 2 \& 6).

\subsubsection{High Temperature, Cool Core Clusters}
\begin{figure}
\includegraphics[width=\columnwidth]{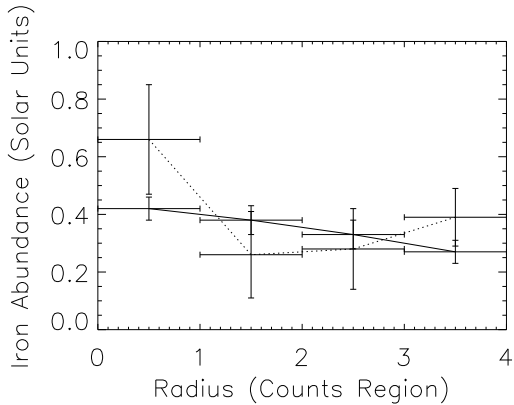}
\caption{\label{Fig2} Iron abundance as a function of radius for
  clusters with high temperatures and short cooling times, separated
  by redshift. The solid line corresponds to subset \# 2 ($z < 0.4$). The dashed 
line corresponds to subset \# 6 ( $z > 0.4$).}
\end{figure}

There are only two clusters in the high redshift subset for clusters
with high temperatures and cool cores (subset \# 6) which is
too small of a sample to be
considered representative of the entire population. However, it
should nevertheless be noted that the average iron abundance profile
for this small subset is entirely consistent point-for-point with the
results of the
larger low redshift sample. As can be seen in Figure 
\ref{Fig2}, the average profile in both cases has a higher central
abundance, with evidence for a decreasing gradient outward. Even
though the average profile for the high redshift sample may not 
represent the population, this small sample
still shows no evidence for evolution in the iron abundance profile
with redshift. 

\subsubsection{Low Temperature, Non-Cool Core Clusters}

\begin{figure}
\includegraphics[width=\columnwidth]{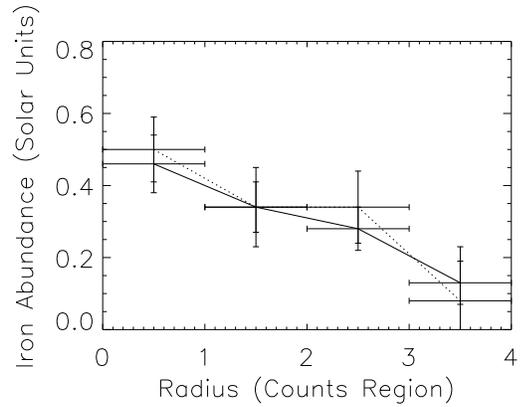}
\caption{\label{Fig3} Iron abundance as a function of radius 
for clusters with low temperatures and long cooling times, separated by redshift.
 The solid line corresponds to subset \# 3 ($z < 0.4$ ).  The dashed line corresponds to subset \# 7 ($z > 0.4$).}
\end{figure}

Subsets \# 3 \& 7 describe the average profiles for low temperature
clusters without cool cores. Although both of these samples are small,
Figure \ref{Fig3} shows that the two profiles are very consistent with
one another. 

\subsubsection{High Temperature, Non-Cool Core Clusters}
Both subsets in this case have a relatively high number of clusters. Again in these
two subsets, the two radial profiles are wholly consistent with one another,
with no reason to believe that the iron abundance in these subsets is
changing with redshift. All of the analysis suggests that the iron
abundance is not evolving with redshift independent of the temperature
and cooling time.

\begin{figure}
\includegraphics[width=\columnwidth]{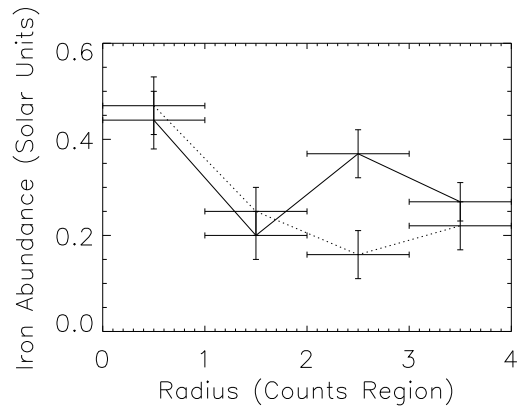}
\caption{\label{Fig4} Iron abundance as a function of radius for
 high temperature clusters with long cooling times, separated by
  redshift. The solid line corresponds to subset \# 4 ( $z < 0.4$). The dashed
line corresponds to subset \# 8 ($z > 0.4$). }
\end{figure}

\section{Discussion} 
\label{disc}
\subsection{The Data Analysis}
Before discussing the radial dependencies of the iron abundance and
the temperature, it is worthwhile to review why and how the data from
the clusters were averaged together in the manner that was done. As
noted in $\S$ 2.3 and 2.4, two different radii/scales were
chosen. The first was the core radius derived from a single $\beta$
model. This radius corresponds directly to the dynamical scale of
the cluster. This is the natural choice for a scale length here, as it
is based on the dynamics of the cluster. 

A second scale length was used to find four radial regions with
approximately equal numbers of net counts.  Although this scale is
independent of the dynamics, it is useful for testing whether the profiles are
sensitive to the scale length chosen.  This "equal counts per enclosed area"
process was chosen for its simplicity in calculation, and for its
ability to increase the significance of measurements in the outer
regions of the clusters.  When the counts and core radii were
converted into a linear projected length scale such as Mpc, they cover
very similar ranges. The main inconsistency occurs in the
outermost regions, where the counts radius extends much further
outward than the core radius. Therefore, overall consistency
between the results derived by the two different methods was
expected. Since the counts radius is so strongly correlated with the
core radius and offers better statistics across each cluster, the 
results of the core radius measurements have been omitted from the text even though
they are wholly consistent with these results. 

Another possible choice of length scale is the virial radius of the
cluster, but the easiest way of estimating r$_{\rm{virial}}$ (or
r$_{200}$) relates directly to the overall temperature
\cite[e.g.,][]{Jones}.  If the virial radius was used in this case, the temperature would be correlated with the radial regions. \rm
Because of the $L_{\rm {x,bol} }-T$ relationship \citep[e.g.][]{Ota} 
and clusters at 
higher redshifts must be on average more luminous to be detected with
a sufficient statistical significance, the scale radius chosen needs to take into account this selection effect. Because the virial radius
is related to the overall temperature, and therefore luminosity, it
is difficult to correct for this selection effect. As an example, there
exist nearby clusters such as Perseus \citep{Perseus} and Coma
\cite[e.g.,][]{Watanabe,DeGrandi, AdamiComa2006} with comparable X-ray
luminosities that exhibit very different temperature and iron
abundance profiles.

As a further check of our analysis, our results were compared with
other published measurements such as \cite{Kotov} for CL 0016+16 (MS 0015.9+1609), and
the temperature/abundance results published here tend to be higher by  $ \sim 1-1.5 \sigma$ than those of 
\cite{Kotov}. The work of \cite{Balestra} shows  how using an updated analysis tool can bring out
difference of this scale. Furthermore, the differences may be due to systematics 
between the \cha and \xmm analysis chains as well as the statistics 
between the two observations themselves. These differences in the core radius measurements are on the order of about 20\% which should not strongly affect the results of the average profiles. The measured profiles of \cite{Kotov} also measure the radial profile in a different energy band than our measurement (0.5-2 keV in \cite{Kotov} versus 0.5-10 keV in this work). \rm Three
other clusters were studied by \cite{Snowden}, and these results are consistent within 10-20\% and usually within 1 standard deviation. An
important caveat is that \cite{Snowden} used the MOS cameras of \xmm,
whereas \cha was used in this study for measuring the temperature of
Abell 1413 and Abell 2204. The systematic differences in
response and sensitivity may be responsible for the
discrepancies between the results of \cite{Snowden} and those listed
here.

The results of the counts profiles results will now be taken at face value, and the physical
interpretation of noticeable variations will be discussed. 

\subsection{Review and Comparison With
Previous Work on the Chemical Evolution of the ICM} \cite{Balestra},
\cite{Maughan2008}, and \cite{Werner} all discuss evidence for the chemical evolution of the ICM.  
Older works \citep[e.g.,][]{Mushotzky1997, Rizza1998} did not report any
statistically significant evidence for chemical evolution with redshift.   The recent work that found
the strongest evidence for evolution is \citeauthor{Balestra}, but this evolution was found only when
data based on all the  $ z < 1 $ clusters in their sample were combined (compare with the smaller previous set in
their Fig. 3). Furthermore, the derived chemical evolution required  the use of the formal statistical best fit
uncertainty instead of the observed dispersions (e.g. their Fig. 14).  In contrast, for a similar
redshift range (below about 0.9),  the results of \citeauthor{Maughan2008} are consistent with no
evolution.

From the theoretical point of view as described in 
\cite{Kapferer}, \cite{Sarazin}, and references therein, 
the chemical evolution of the ICM  is a complex combination of effects due to 
cluster merging,
infall of enriched material, galactic ram pressure stripping, galactic winds, and other
possible causes.
For the relaxed clusters considered in this paper \cite{Kapferer} predicted no observable chemical
evolution  from redshift 0.15 to 0.9.

Our paper shows how it is important to compare like clusters and 
that there is
room for further theoretical studies to understand the apparent lack of evolution
of the radial dependence of the enriched material of the ICM, and at the 
same time
induce heating via infall of metal rich gas clouds.
 
\rm 
\subsection{Scenarios for heating and iron enrichment}
\label{scenarios}
The data presented here are
consistent with no evolution in radial profiles of
iron abundance and temperature. The modeling of this should be consistent with the concept of hierarchical
formation of structure in the universe \cite[e.g.][and references
therein]{Gao2008}. Because the iron abundance does not appear to be
strongly correlated to the overall temperature, gravitational infall
is the most likely scenario for heating clusters beyond $\simeq $ 6.5
keV. This infall must take
place in such a manner as not to change the radial profile of iron abundance 
or temperatures. For non-cooling
clusters this means the clusters somehow have no measurable temperature
gradients inside about 500 kpc but maintain their abundance
gradients while (presumably)  increasing in temperature in a hierarchical 
growth
model.  In order to demonstrate why infall is the preferred major
energy input, the process of heating has to be considered in some detail. 

An estimate of the amount of near solar abundance material that is
added along with the energy to boost the $<\!\!kT\!\!>$ from 4 keV
to 8 keV is made now. We assume typical $L_{X}-T$
\citep{Ota,Ettori2004a,Stanek2006} and $L_{X}-M$
\citep[e.g.][]{Rykoff2008,Stanek2006} relations apply
to the clusters studied here. Under these assumptions, if the temperature
increases by
 a factor of 2  then the luminosity increases by at least a factor of 4 and the mass
by a factor of 3, the amount of mass added to the clusters being on the
order of $3 \times 10^{14} M_{\odot}$.  Since galaxies are thought to
make up only about 1/5 of the baryonic mass in clusters
\citep[see][and references there in]{Lowenstein2006}, it is implausible
that this mass and additional gravitational energy is added by normal
galaxies. 
This mass and gravitational energy must come instead
from the infall of atypical galaxies that have gas masses that greatly
exceed their stellar masses or gas clouds that never formed into
galaxies. Damped Lyman $\alpha$ absorbers (DLAs) have metallicities
that are near solar at  $z$ of 1 or higher \citep[][and references
therein]{Meiring2007} which makes 
the
ability to add both mass and metals via infalling clouds plausible. It is beyond the
scope of this work, though, to carry out detailed calculations of
this infall scenario. However, an implication of the estimates made
here is that
sight lines on the outskirts of clusters should show the presence of
DLAs at the cluster redshift.  Furthermore, if mass is added with
energy as implied by the $L-T$ and $L-M$ relationships used here, this
rules out processes which might provide energy and metals but not
significant amounts of additional mass, such as
SN or AGNs.  

Instead, suppose that the $L-M$ relationship used here doesn't apply to
these 
high $z$ clusters. In this case, supernovae
would still not be a plausible explanation.  
This is because the energy input of approximately $10^{64}$ ergs would 
require an 
unreasonably high
number of SNe when  all the available energy is transferred to the ICM
\citep{Conroy2008}. 
For example, suppose there are 1,000 galaxies per cluster. This
translates into $10^{10}$ SNe per galaxy or $10^{1}$ SNe/year for
$10^9$ years. It is also implausible that a central AGN could provide
this 
much heat
($10^{64}$ ergs), as this heat would require
$10^{55}$ erg/year deposited for 1 Gy, or $10^{47}$ erg s$^{-1}$
minimum
 energy generation
assuming 100\% efficiency in transfering energy to heat. 
Since the magnitude of heating seems beyond what central AGN activity
could
 provide 
an infall scenario is a more likely explanation.

\subsection{Scenarios for Mixing}
Since there is no evidence for evolution in every type of cluster
considered in this study (as seen in the figures), it
will be assumed for the sake of discussion that clusters are not
mixing over this range of redshifts.

Beginning with the cool core clusters in Fig. \ref{Fig1}, it is seen
that they exhibit almost no signs of evolution in their iron abundance profiles
from the 0.4-0.9 redshift bin up to the present day,
even though there is a very clear gradient at all redshifts. This 
suggests two possible scenarios: either enrichment and mixing both
exist in such a
way that neither is dominant, or neither process
occurs. If neither process occurs, the unchanging iron abundance
profile can be explained as being due to the average cluster 
galaxy having lost most of its
gas by $z \sim 0.8$. In this case the galaxies have no gas to stir up and mix
the ICM. An absence of gas in cluster galaxies would also manifest
itself in a low, non-evolving star formation rate\citep{Homeier2005}. A low star formation rate occurs if  ram pressure stripping has removed the majority of the gas from most of
the cluster galaxies before redshift $z = 0.8$ within 500 kpc of the
cluster center, well within the regions observed
in this study. The suggestion that most gas is stripped from
galaxies by $z \simeq 0.4$ was also seen by \cite{Butcher, Dressler}.

As the gradient in iron abundance has remained
constant with redshift, there is also no evidence for growth in the
gradient.  Substantial changes in the abundance gradient would be expected if there was continuous activity from the central AGN . Therefore, energy input
to the ICM beyond gravitational
infall \citep{Bode2007} must not have caused
appreciable iron enrichment. 

For non-cool core and high temperature clusters, there is less
statistical evidence to suggest that the iron abundance decreases with
radius in the same manner as low temperature cool-core
clusters. However, as noted above, it is assumed there is no
evolution in the radial profiles of the ICM with redshift.  In this case, 
galactic motion within the cluster does not result in ram pressure
stripping, one of the key processes for iron enrichment 
(and possibly mixing) in the
ICM. Without ram pressure stripping from galaxies to enrich or mix the
ICM, the iron abundance (and iron abundance profile) of the ICM remains static 
over long times. 

\section{Summary and Conclusions} 

Average radial profiles of the temperature and iron abundance of X-ray bright clusters
in the redshift range from about 0.14 to 0.9 were calculated. We find no evidence for evolution within similar sets of clusters.
The total element abundance remains constant with
decreasing $z$ and the gradients remain approximately the same.  These
observations suggest that gravitational infall is the dominant mechanism for heating of the ICM, and
also that ram pressure stripping does not substantially change the ICM
out to large redshifts. 
If clusters experience a
significant (factor of 2 or more) amount of gravitational mass infall, 
then the iron abundance profile
remains constant while the overall temperature increases.  This
infall needs to simultaneously keep the profile (within error of about 20\%) 
isothermal while also
maintaining the same iron abundance profile.  Mass and energy
considerations suggest infall of metal rich gas, but how the process
takes place needs further theoretical study.  These same considerations
of mass and energy suggest that although central AGN activity is
directly related to cooling cores, the central AGN does not produce 
significant amounts of metal rich gas and energy to the ICM beyond
its local environment of about 50-100 kpc. 

Other theoretical challenges to consider include how clusters 
formed in the first place with both abundance gradients and isothermal 
temperature profiles. 
Work also needs to be done on how the frequency and scale of
merger activity \citep[e.g.][]{Wik2008} changes the abundance and 
temperature gradients of clusters.  Another potentially 
important aspect of 
modeling galaxy clusters theoretically appears to be an accurate
modeling of the local environment surrounding the cluster, 
as the infall of material initially outside of the cluster could play an 
important role in its evolution.

From an observational point of view it will be interesting to measure
temperature and abundance
gradients for clusters at $z$ greater than 0.9 as well as make measurements
of DLAs by means of QSOs whose sight lines are at $\simless 1$ Mpc  distances
from the centers of rich clusters. 
These measurements may help determine more conclusively
whether 
galaxy clusters are evolving 
in the ways described here 
and what kind of material 
might be falling in to cause the highest temperature clusters of galaxies.

\acknowledgement
This research has made use of the NASA/IPAC Extragalactic Database
(NED) which is operated by the Jet Propulsion Laboratory, California
Institute of Technology, under contract with the National Aeronautics
and Space Administration.

We would like to thank the Max Planck Institut f\"{u}r Kernphysik in
Heidelberg, Germany and Le Laboratoire d'Astrophysique de Marseille in
Marseille, France for their hospitality. We would also like to
thank A. Kathy Romer and Kivanc Sabirli for providing processed XMM
data used in the early stages of this analysis. We would also like to
thank Wilifried Domainko for his review and suggestions for
improvement. We thank the referee for many insightful comments that
greatly improved the manuscript.  Finally, we would like to acknowledge
support from NASA Illinois Space Grant NGT5-40073.

\bibliography{aa_2008_09809}
\bibliographystyle{aa}

\end{document}